\documentclass[english,twoside,a4paper,10pt]{article}

\usepackage[latin1]{inputenc}
\usepackage[T1]{fontenc}
\usepackage{amsmath}
\usepackage{amsfonts}
\usepackage{graphicx}
\usepackage{subfigure}
\usepackage{a4wide}
\usepackage{amssymb}
\usepackage{fancyhdr}
\usepackage{mathrsfs}
\usepackage{amssymb}
\usepackage[toc,page]{appendix}
\usepackage{xcolor}
\usepackage{float}
\usepackage{cite}

\def\ba{\begin{eqnarray}}
\def\ea{\end{eqnarray}}

\def\be{\begin{equation}}
\def\ee{\end{equation}}

\begin{document}

\title{A class of static spherically symmetric solutions in $f(T)$-gravity}

\author{ 
Marco Calz\'a$^1$\footnote{E-mail address: mc@student.uc.pt
},\,\,\,
Lorenzo Sebastiani $^{2}$\footnote{E-mail address:lorenzo.sebastiani@unitn.it}
\\
\\
\begin{small}
$^1$ CFisUC, Departamento de Fisica, Universidade de Coimbra, 3004-516 Coimbra,
Portugal
\end{small}\\
\begin{small}
$^2$ Dipartimento di Fisica, Universit\`a di Trento, Via Sommarive 14, 38123 Povo (TN), Italy
\end{small}\\
}

\date{}

\maketitle

\abstract{We study a class of static spherically symmetric vacuum solutions in modified teleparallel gravity solving the field equations for a specific model Ansatz, requiring the torsion scalar $T$ to be constant. We discuss the models falling in this class. After some general considerations, we provide and investigate local solutions in the form of black holes and traversable wormholes as well as configurations that can match the anomalous rotation curves of galaxies.

\section{Introduction}

General Relativity (GR) has been tested for more than a century and it represents the best theory of gravity available so far. Nonetheless, it deals with problems in its infrared (IR) and ultraviolet (UV) aspects and cannot be regarded as the final theory of gravity. In the IR scales, at cosmological and astrophysical scales, dark components need to be invoked to describe the observed scenarios and at the UV scales, GR presents challenges in its quantization due to the fact that it is not a renormalizable theory.
This scenario is a quest to go beyond GR, as it has been widely done in recent decades.

A building block at the base of the current interpretation of gravity is the Equivalence Principle (EP), an empirical and well-measured fact forcing gravity to be a geometrical theory. More specifically, the EP together with the requirement that different observers need to describe nature with the same physical laws 
led to the construction of the theory of gravity as a metric-affine theory.

In this context, GR is a geometric theory of gravity describing space-time as a  manifold in which the affine connection is metric-compatible and
torsion-free and it is given by the Levi-Civita connection totally defined by the metric. Equivalently, one can say that GR is based on a Riemannian manifold on which the fundamental quantity is the scalar Ricci curvature.

This choice although fruitful is arbitrary since three fundamental geometrical quantities characterize a general metric-affine theory. Namely, the curvature $R$, the torsion $T$, and the non-metricity $Q$ \cite{Iosifidis:2021pta,BeltranJimenez:2019esp}.

Metric-affine manifold can be classified according to the number of non-vanishing scalars defining its connection. If none of $R$, $T$, or $Q$ is vanishing the manifold is the most general possible, and it is just dubbed as matric-affine.
Sub-classes are the torsion-free ($T = 0$), Riemann-Cartan  ($Q = 0$), and teleparallel ($R = 0$) manifolds. Moreover, further subsets are obtained if two scalars vanish simultaneously as in the case of Riemannian ($T = Q = 0$), Weitzenb\"ock or teleparallel ($R = Q = 0$),
and symmetric teleparallel ($R = T = 0$) manifolds. Finally, when the three quantities vanish together we are in the trivial subset of a merely Minkowskian manifold.

It is now well understood that GR is just a vertex of a geometrical trinity of dynamically equivalent gravity theories in which the other vertexes are the Teleparallel Equivalent to General Relativity (TEGR) \cite{Springer,Maluf:2013gaa, moltoaltro, moltoaltro1, moltoaltro2, moltoaltro3, moltoaltro4, moltoaltro5, moltoaltro6, moltoaltro7, moltoaltro8},  and the Symmetric Teleparallel Equivalent of General Relativity (STEGR) \cite{Nester:1998mp,Adak:2004uh,Adak:2005cd,Adak:2008gd,Mol:2014ooa,BeltranJimenez:2017tkd,BeltranJimenez:2018vdo,Gakis:2019rdd}. The former is characterized by vanishing curvature and non-metricity and the connection reduces to the Weitzenb\"ock connection, while in the latter, the curvature and the torsion vanish. Both the equivalent theories are defined by a  lagrangian density coincident with the respective scalars $T$ and $Q$ (similarly to the lagrangian density of GR characterized by the scalar curvature $R$). These equivalent formulations have recently been compared and reviewed in Refs. 
\cite{Jarv:2018bgs,Capozziello,Heisenberg:2018vsk}.

Taking the non-relativistic limit of those theories one can show that the equivalency is preserved in the non-relativistic limit and it is possible to build equivalent non-relativistic geometric models of gravity \cite{Wolf:2023rad}, namely Newtonian Gravity theory, Newton-Cartan theory, and Symmetric Newton-Cartan theory.
On the other hand, extended theories of gravity $f(R)$ \cite{DeFelice:2010aj,Sotiriou:2008rp}, $f(T)$ \cite{Cai:2015emx,Bahamonde:2021gfp, fT1, fT12, fT13, fT41, fT42, SariRep, mioultimo, Z1, Z2, Z3, Z4,Briffa:2023ozo} , and $f(Q)$ \cite{BeltranJimenez:2017tkd,BeltranJimenez:2018vdo,Zhao:2021zab,Lazkoz:2019sjl,Mandal:2020lyq, Mandal,Capozziello:2022tvv,Capozziello:2022wgl,Hu:2022anq}where the lagrangians are given by generic functions of $R\,,T$ and $Q$, respectively, do not generally display the same equivalence. The main cause of this lack of equivalence dwells in the fact that the Equations of Motion (EoMs) of $f(R)$-gravity are fourth-order, while the other two theories are characterized by second-order EoMs. Nonetheless, it was recently shown \cite{Capozziello:2023vne} that by taking into account the boundary terms of TEGR and STEGR ($\tilde B$ and $B$ respectively) it is possible to promote the extended theories from the second to the fourth order by considering $f(\tilde B -T)$ and $f(Q-B)$ and restore the dynamical equivalence.

The direct observation of the black hole (BH) shadows at the center of the Milky Way \cite{EventHorizonTelescope:2022xnr} and of M87 \cite{EventHorizonTelescope:2019dse}
by Event Horizon Telescope and the very many LIGO/Virgo/KAGRA Collaboration merging events \cite{LIGOScientific:2016aoc} motivate the research of local solutions in the form of BHs, Wormholes (WHs), or any ultra-compact object, the simplest description of which is provided by static spherically symmetric (SSS) geometries. Moreover, SSS geometries may also take into account another well-studied problem in experimental physics: the anomalous rotation of spiral galaxies. In those respect, SSS local solutions have been already explored in $f(R)$-gravity \cite{Multamaki:2006zb,delaCruz-Dombriz:2009pzc,Kobayashi:2008tq,Upadhye:2009kt,Calza:2018ohl}, in $f(T)$-gravity \cite{Aftergood:2014wla,Bahamonde:2019jkf,Ilijic:2018ulf,LinfT,DeBenedictis:2016aze,Beau,Krssak:2015oua,Krssak:2015rqa} and lately also in $f(Q)$-gravity \cite{Lin,Hassan:2022jgn,Tayde:2022lxd,Sokoliuk:2022efj,Banerjee:2021mqk,Hassan:2022hcb,Maurya:2022wwa,Parsaei:2022wnu,Wang:2021zaz,Mandal:2021qhx,DAmbrosio:2021zpm,Bahamonde:2022esv,Bahamonde:2022zgj,Calza:2022mwt}.

In this paper we will investigate SSS solutions in the framework of $f(T)$-gravity and respecting a particular Ansatz that solves on shell the EoMs.

In Refs. \cite{Calza:2018ohl,Calza:2022mwt} we analyzed SSS space-time vacuum solutions in the framework of $f(R)$ and $f(Q)$ gravity. We considered an Ansatz solving identically the EoMs of special classes of models when the respective scalar quantity is constant (eventually vanishing). Here, we conclude the study and exploit once again the Ansatz to find SSS vacuum solutions in 
$f(T)$-gravity models respecting our Ansatz.  
In this way, the EoMs are automatically satisfied in vacuum without solving them explicitly and a wide class of new solutions can be found. We discuss several possibilities and focus on solutions capable of describing BHs, WHs, and rotational curves of galaxies.

This paper is organized as follows. In Sec. {\bf 2} we revisit the formalism of $f(T)$-gravity. In Sec. {\bf 3} we introduce an Ansatz selecting a class of $f(T)$-models with exact SSS solutions for constant torsion scalar. In this respect, we also provide some examples of models. Sec. {\bf 4}, Sec. {\bf 5} and Sec. {\bf 6} are devoted to the study of vacuum solutions describing BHs, WHs and resembling the profile of rotation curves of galaxies, respectively. 
Conclusions and final remarks are given in Sec. {\bf 7}.

In this work, we use units of $k_B = c = \hbar = 1$ and we denote the gravitational constant $\kappa^2 = 8\pi G_N$.

\section{$f(T)$-theories of gravity}

We describe some general features of $f(T)$-theories of gravity. Teleparallel gravity is a variant of Riemann-Cartan geometry where a spin connection is present.
In TEGR one may use the notion of a proper frame in which the spin connection is vanishing, since the term in the torsion scalar depending on the spin connection can be rewritten as a total derivative, and disappears from the equations of motion derived from a Lagrangian which is still invariant under local Lorentz transformations \cite{Kr}. However, in $f(T)$-modified theories of gravity, the variation of such term will be in general not vanishing
and this choice leads to a manifest breaking of the local Lorentz invariance and makes the theory frame-dependent. In this section we use the covariant formalism presented in Ref. \cite{Krssak:2015oua}, using both the tetrad $h^a_\mu$ and the spin connection $\omega^a_{b \mu}$ as dynamical variables
in order to avoid the violation of local Lorentz invariance. For a comprehensive discussion of the covariant formulation of $f(T)$-gravity we recommend to the careful reader Refs. \cite{Krssak:2015oua,Krssak:2015rqa}.

The tetrad is a set of four orthonormal vectors representing a reference frame for the physical observer and is related to the metric tensor through
\begin{equation}\label{tetrad}
   g_{\mu \nu}= \eta_{a b} h^a_\mu h^b_\nu\,,
\end{equation}
while for the spin connection, we have
\begin{equation}\label{affine}
    \omega^a_{b \mu}= \tilde \omega^a_{b \mu} + K^a_{ b \mu} \,.
\end{equation}
Here, $\eta_{a b} = diag(-1,1,1,1)$, 
and $K^a_{ b \mu} $ takes the name of contortion tensor,
\begin{equation}\label{contortion}
   K^a_{b \mu}= \frac{1}{2}  T^a_{\;\;b \mu}+ T^{\;\;\;a}_{(b \;\; \nu)}\,,
\end{equation}
where $T^a_{\;\;b \mu}$ is the torsion tensor defined as
\begin{equation}\label{torsion}
   T^a_{\;\;b \mu}= \partial_\mu h^a_\nu -\partial_\nu h^a_\mu + \omega^a_{b \mu} h^b_\nu - \omega^a_{b \nu} h^b_\mu\,.
\end{equation}
Analogously to what is done for curvature where the Ricci scalar $R$ takes into account an amount of curvature, one can define a scalar quantity taking into account the amount of torsion, namely the torsion scalar
\begin{equation}\label{scalartorsion}
   T=T^a_{\mu \nu}S^{\mu \nu}_a\,,
\end{equation}
where.
\begin{equation}\label{dual}
   S^{\mu \nu}_a=K^{\mu \nu}_a-h^\nu_a T^{\alpha \mu}_\alpha + h^\mu_a T^{\alpha \nu}_\alpha\,.
\end{equation}
The general Lagrangian density of modified teleparallel gravity is given by,
\begin{equation}\label{Ldens}
   \mathcal{L}=\frac{h}{4 \kappa^2} f(T)\,,
\end{equation}
where $h=\text{det}( h^a_\mu$), and $f(T)$ is an arbitrary function of the torsion scalar only.

The field equations are derived through variations with respect to the tetrad and read
\begin{equation}\label{fieldeq}
   h^{-1}f_T \partial_\nu(h S^{\mu \nu}_a)+ f_{TT} S^{\mu \nu}_a \partial_\nu T - f_T T^b_{\nu a} S_b^{\nu \mu} + f_T  \omega^b_{a \nu} S_b^{\nu \mu}+ \frac{1}{4}f h^\mu_a= \kappa \mathcal{T}^\mu_{a}\,,
\end{equation}
where $f\equiv f(T)$, $f_T=\frac{d f(T)}{dT}$,  $f_{TT}=\frac{d^2 f(T)}{dT^2}$, and
\begin{equation}
   \mathcal{T}^\mu_{a}= \frac{1}{h} \frac{\delta \mathcal{L}_m}{\delta h^\mu_a}\,,\label{stresstensor}
\end{equation}
with $\mathcal{L}_m$ the usual Lagrangian density of matter.

\section{The model Ansatz \label{ModelAn}}

We aim to focus on vacuum solutions, namely solutions with $\mathcal{T}^\mu_{a}=0$. In this way the simple assumption
\begin{equation}
f(T_0)=f_T(T_0)=0\,,
\label{Ansatz}
\end{equation}
where $T=T_0$ is a constant value (eventually vanishing) of the torsion scalar,
 automatically satisfies the equations of motion (\ref{fieldeq})--(\ref{stresstensor}). 

This Ansatz does not force the function $f(T)$ to assume any specific functional shape which then is not uniquely determined. Therefore, the Ansatz spans different possible choices of specific modified teleparallel gravity of physical interest.
For example, polynomial models of the type 
\begin{equation}
f(T)=\gamma(T-T_0)^n\,,
\quad n\geq 2\,,
\end{equation}
where $\gamma$ is a dimensional constant,
fall in this class of theories. We can write,
\begin{eqnarray}
f(T)&=&\gamma \sum_{k=0}^{n}
\frac{n!}{k!(n-k)!}T^{n-k}(-T_0)^k\,,
\nonumber \\ 
&=&\gamma \frac{n!}{(n-1)!}(-T_0)^{n-1} T
+\gamma (-T_0)^n+
\gamma
\sum_{k=0}^{n-2}
\frac{n!}{k!(n-k)!}T^{n-k}(-T_0)^k\,,
\end{eqnarray}
and by posing $\gamma\frac{n!}{(n-1)!}(-T_0)^{n-1}=1$ at the leading order, apart from a cosmological term, we find a power-law correction to TEGR  which is relevant only for large values of the scalar torsion $T$.

Moreover, some applications to dark energy phenomenology can be obtained by considering the model,
\begin{equation}
f(T)=T+2\Lambda\left(1-\text{e}^{\frac{T}{2\Lambda}}\right)\,,    
\end{equation}
where $\Lambda$ is the cosmological constant, which is equivalent to the so-called one-step models of $f(R)$-gravity \cite{onestep, onestep2,onestep3, onestep7,onestep8,onestep9,onestep10}.

In this context, we are interested in SSS solutions whose line element reads
\begin{equation}
ds^2=-h(r)dt^2+\frac{1}{g(r)} dr^2+r^2 d\Omega^2 \,,\label{metric0}
\end{equation}
where $h(r)\,,g(r)$ are generic functions of the radial coordinate $r$ and $d\Omega^2=\left(d\theta^2+\sin\theta d\phi^2\right)$ represents the metric of the two-dimensional sphere. Thus, the scalar torsion is given by \cite{Beau}
\begin{equation}
T=\frac{4g(r)}{r^2}\left(\frac{1}{\sqrt{g(r)}}-1\right)
\left(\frac{1}{\sqrt{g(r)}}-1-\frac{r h'(r)}{h(r)}\right)\,.
\end{equation}
The prime index corresponds to the derivative with respect to $r$.
We will look for solutions where
\begin{equation}
\frac{4g(r)}{r^2}\left(\frac{1}{\sqrt{g(r)}}-1\right)
\left(\frac{1}{\sqrt{g(r)}}-1-\frac{r h'(r)}{h(r)}\right)=T_0\,,\label{T0}
\end{equation}
with $T_0$ constant,
which are solutions of the class of models previously discussed. As a special case, when $T_0=0$, 
we can choose $g(r)=1$ (see \S\ref{rotation}) or, alternatively, we have to impose
the following condition
\begin{equation}
\left(\frac{1}{\sqrt{g(r)}}-1-\frac{r h'(r)}{h(r)}\right)=0\,.\label{T0=0}
\end{equation}

\section{Black hole solutions}

In this section
we are interested in black hole  solutions in the form of (\ref{metric0}) and satisfying Eq. (\ref{T0}) for some value of $T_0$, which are exact solutions of $f(T)$-modified gravity models introduced in \S \ref{ModelAn} for which condition (\ref{Ansatz}) holds true. 

Let us start by considering the Schwarzshild gauge with $h(r)=g(r)$, namely
\begin{equation}
ds^2=-g(r)dt^2+\frac{1}{g(r)} dr^2+r^2 d\Omega^2 \,.
\end{equation}
Thus, the zeros of $g(r)$ correspond to BH horizons as soon as $g(r)>0$ in order to have a positive surface gravity.

Therefore, Eq. (\ref{T0}) leads to
\begin{equation}
    g(r)=h(r)=1-\frac{c_1}{r}+\frac{r^2 T_0}{12}\pm 2\sqrt{-\frac{c_1}{r}+\frac{r^2 T_0}{12}}\,,\label{sol1}
\end{equation}
where $c_1$ is an integration constant, while $T_0$ is fixed by the model. If $T_0>0$ 
and $c_1>0$ a minimal length scale for the radial coordinate appears since $r\geq r_0=\left(\frac{12c_1}{T_0}\right)^\frac{1}{3}$ and the singularity at $r=0$ is avoided. 
 On the other side, when $T_0<0$, $c_1$ has to be negative, namely $c_1<0$, and the metric is still free of central singularity. In all these cases one expects that $r_0$ is a Plankian size, namely $c_1/T_0\ll 1$.

Solution (\ref{sol1}) can be written as
\begin{equation}
g(r)=h(r)=\left(1\pm X(r)\right)^2\,,
\quad X(r)=\sqrt{-\frac{c_1}{r}+\frac{r^2 T_0}{12}}\,,
\end{equation}
and the interesting case for the formation of an event horizon is the one with the negative sign for which $g(r)$ and $h(r)$ vanish when $X(r)=1$, namely
\begin{equation}
r^3 T_0-12r-12c_1=0\,.
\end{equation}
This equation may possess one positive root when $T_0>0$ and   $c_1>0$ or when $T_0<0$ and $c_1<0$
and two positive roots when
$T_0>0$ and
$c_1<0$ 
(Descartes'rule). In the second case, we are in the presence of a black hole with an additional internal Cauchy horizon (for the issues related to the instability of the Cauchy horizon see Refs. \cite{ins1, ins2}).
In any case, $g(r)$ and therefore $h(r)$ are positively defined, eventually vanishing, and they never take negative values. Thus, the solution does not show the usual inversion of the Killing vectors through the horizon.
Moreover, $g'(r)=0$ on the BH horizon when $X(r)=1$ and the Hayward surface gravity $\kappa_H$ \cite{Hay} vanishes therefore the temperature associated to the BH horizon $T=\frac{\kappa_H}{2\pi}=\frac{g'(r)}{4\pi}$ is null as in an extremal black hole \cite{Ghosh}.
Finally, when $T_0\neq 0$, in the external region far from the horizon, the solution turns out to coincide with de Sitter (dS) or an Anti-De Sitter (AdS) space-time, depending on the sign of $T_0$.

 To better understand the causal structure of the solution, let us consider the change of coordinates,
\begin{equation}
ds^2=-h(r)du^2- 2\sqrt{\frac{h(r)}{g(r)}}du dr+r^2 d\Omega^2 \,,\quad u=t+r^*\,,\nonumber
\end{equation}
\begin{equation}
ds^2=-h(r)dv^2+2\sqrt{\frac{h(r)}{g(r)}}dv dr+r^2 d\Omega^2 \,,\quad v=t-r^*\,,
\end{equation}
where the null coordinates $u\,,v$ are determined by the tortoise coordinate $r^*$ which is given by,
\begin{equation}
dr^*=\frac{dr}{\sqrt{h(r)g(r)}}\,.
\end{equation}
We recall that the light cone is defined as the surface for which the incoming and outgoing null geodesic $du$ and $dv$ vanish, 
namely when $dv=0$ and $dv=-2/\sqrt{h(r)g(r)}dr$ or when  $du=0$ and  $du=2/\sqrt{h(r)g(r)}dr$.
Thus, introducing the new coordinate $t^*=u-r$ or $t^*=v-r$, one can describe the light-cone with
\begin{equation}\label{lightcone}
    \frac{dt^*}{dr}=-1 \;\;\;\;\;\text{and}\;\;\;\;\; \frac{dt^*}{dr}= \pm \left( \frac{2}{\sqrt{h(r)g(r)}}-1\right)\,.
\end{equation}
To move forward, we can investigate the special case $T_0=0$ in (\ref{sol1}), such that we deal with an asymptotically Minkowskian solution having
\begin{equation}
    g(r)=h(r)=1-\frac{c_1}{r}\pm
    2\sqrt{-\frac{c_1}{r}}=\left(1 \pm \sqrt{-\frac{c_1}{r}}\right)^2
    \,,
\end{equation}
with $c_1<0$.
Once again, the BH solution is the one with a negative sign and the event horizon is located at
\begin{equation}
r=-c_1\,.
\end{equation}
As we already discussed, the metric functions $g(r)$ and $h(r)$ are positively defined inside and outside the horizon, and integrating the two equations in (\ref{lightcone}) one has,
\begin{equation}
  t^*_1=-r+k_1\,,\nonumber
\end{equation}
\begin{equation}
\begin{split}
  t^*_2=&r-6 c_1 \left(\log \left(\left|1+ \sqrt{\frac{r}{c_1}}\right| \right)-\log \left(\left| 1-\sqrt{\frac{r}{c_1}}\right| \right)- \log (| c_1-r| )\right)\\
  &+\frac{-8 \sqrt{c_1 r^3}+12 \sqrt{c_1{}^3 r}+4 c_1{}^2}{c_1-r}+k_2\,,
  \end{split}
\end{equation}
where $k_{1,2}$ are arbitrary constants. Now, the intersection of the functions $t^*_1\equiv t^*_1(r)$ and $t_2^*\equiv t^*_2(r)$ define the structure of the light cone in the $(t^*,r)$-plane. This set of coordinates displays an outstanding advantage with respect to the usual set of coordinates $(t,r)$: the horizon singularity is removed and the light cone does not degenerate into a straight line. We notice that in the far away region we have $t^*_1=-r+k_1$ and $t^*_2 \sim r+k_2$, therefore for different values of $k_{1,2}$ the light cone is described by lines with $45^\circ$ tilt in the $(t^*,r)$-plane with the future cone pointing upward toward positive values of $t^*$, as one expects for Minkowski space-time. 

To better grasp the causal structure of the solution approaching the horizon, we look at Fig. 1 (a) in which it is reported $t^*_1(r)$ for different values of $k_1$ (solid blue lines) and  $t^*_2(r)$ for $k_2=0$ (solid red line), respectively, and, for the sake of simplicity and without loss of generality, we took $c_1=1$. 
The red dotted line at  $r=1$ corresponds to the horizon and there $t^*_2(r)$ diverges and changes its sign crossing the horizon. Coming from radial infinity and approaching the horizon, the light cone tilts on the left and then the right-hand side of the light cone becomes vertical at the horizon.
However, differently to what happens in the Schwarzschild solution, where, inside the horizon, the light cone rotates of $90^\circ$ at $r=0$, here, the causal structure re-approach its configuration at $45^\circ$ for a finite and non-vanishing radial value (see Fig. 1 (b) ) and therefore tilts on the right side.
At $r=0$ the red and blue lines coincide and in that point, there exists no space-like region, namely, in that point space-time is Euclidean and can only be reached by light rays.
Such a BH still possesses an unremovable singularity at $r=0$, but it cures the unphysical feature of having all the matter and radiation accumulated in one point with infinite density.
\begin{figure}[h!]\label{fi1}
    \centering
    \begin{minipage}{0.44\textwidth}
        \centering
    \includegraphics[width=0.99\textwidth]{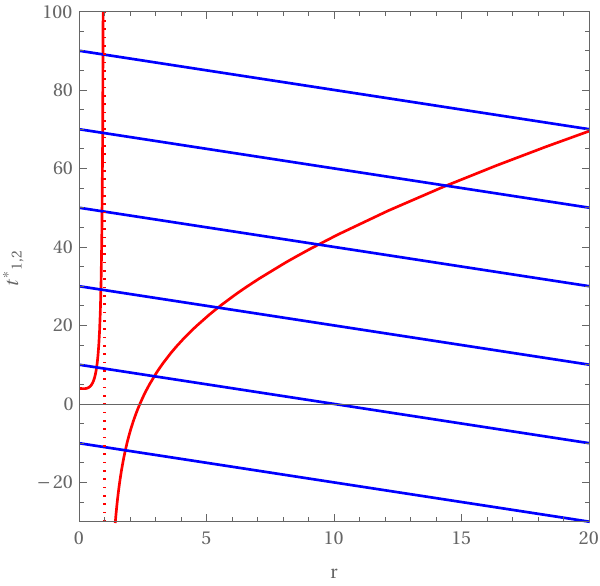}
        \text{(a)}
    \end{minipage}\hfill
    \begin{minipage}{0.44\textwidth}
        \centering
    \includegraphics[width=0.99\textwidth]{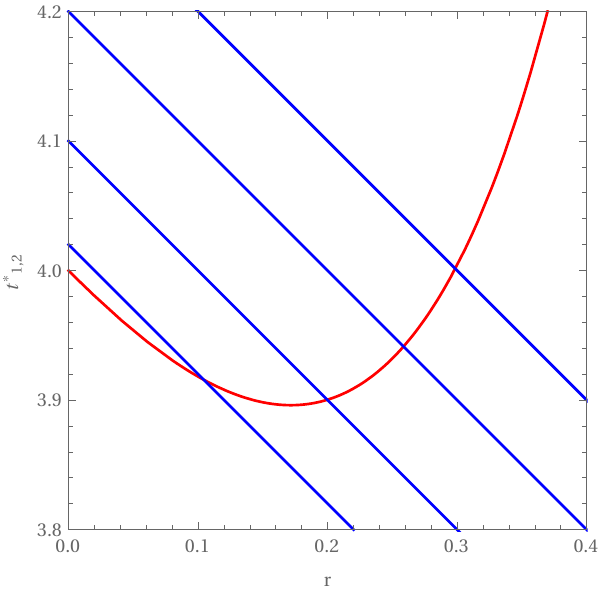}
        \text{(b)}
    \end{minipage}\hfill
    \caption{Here we considered $c_1=1$ and report $t^*_1(r)$ for different values of $k_1$ (solid blue lines) and  $t^*_2(r)$ at $k_2=0$ (solid red line). The red dotted line at  $r=1$ corresponds to the horizon and there $t^*_2(r)$ has an asymptotic behavior and changes sign crossing the horizon.  Fig 1 (b) is a zoom of the same plot regarding the internal structure of the BH. }
\end{figure}
As a consequence, the structure internal to the event horizon is causally disconnected from the external world, no particle can escape and no particle can reach the point $r=0$.\\
\\
Some other BH solutions that lead to $T=T_0$ can be obtained by posing $g(r)=h(r)^2$ in (\ref{metric0}). Thanks to this choice Eq. (\ref{T0}) simply reads,
\begin{equation}
\frac{4}{r^2}\left(1-h(r)\right)
\left(
1-h(r)-rh'(r)
\right)=T_0\,,
\end{equation}
whose solutions are,
\begin{equation}
    h(r)= 1 \pm \frac{\sqrt{r^4 \frac{T_0}{2}-c_1}}{2 r}\,,\label{h2}
\end{equation}
with $c_1$ integration constant. The BH solutions correspond to the ones with the minus sign.
Moreover, if $T_0>0$ and $c_1>0$ or 
$T_0<0$ and therefore $c_1<0$
we have $r>r_0=\left(\frac{8c_1}{T_0}\right)^\frac{1}{4}$ and the central singularity at $r=0$ is avoided. 

If $T_0=0$ and $c_1<0$ the solution 
$h(r)=\sqrt{g(r)}= 1 \pm \frac{\sqrt{- c_1}}{2 r}$ is asymptotic flat and one finds
\begin{equation}
h(r)=1\pm\frac{c_0}{r}\,,
\end{equation}
where $c_0=\frac{\sqrt{- c_1}}{2}$ is a positive constant. The solution with the minus sign describes a black hole whose event horizon is located at $r=c_0$.

We also can generalize the previous solution by taking $h(r)=g(r)^z$, with $z\neq 0\,, -\frac{1}{2}$ a generic number. Then solving Eq. (\ref{T0}) we get,
\begin{equation}
    g(r)=h(r)^\frac{1}{z}=1-\frac{c_1 r^{-1/z}}{2 z}+\frac{r^2 T_0}{4 (2 z+1)} \pm\frac{r^{-2/z} \sqrt{r^{3/z} \left(z \frac{T_0}{2} r^{\frac{1}{z}+2}-2 c_1 z-c_1\right)}}{\sqrt{z^2+\frac{z}{2}}}\,,
\end{equation}
with $c_1$ constant, and various BH configurations for different values of $T_0$ are allowed. Note that for $z=1/2$ we recover solution (\ref{h2}) with $c_1\rightarrow 4c_1$.

\section{Wormhole solutions}

In this section we will investigate exact wormhole solutions \cite{Morris} which lead to a constant torsion scalar $T=T_0$  and which belong to the class of models satisfying the Ansatz in (\ref{Ansatz}).

It is worth recalling that in GR, one can realize a traversable wormhole only by invoking the presence of an exotic matter source that violates the null energy condition \cite{WH,WH2,WH3,WH4,WH5,WH6,WH7,WH8,WH9,WH10,WH11,Calza:2022szy}. However, in the framework of a modified theory,  
wormholes may be realized as vacuum solutions, 
since the unpleasant role of the anti-gravitational matter could be played by the modification of gravity itself \cite{WHMG, WHMG2, WHMG3,WHMG4}.

By making the choice
$h(r)=\text{e}^{2\Phi(r)}$, the line element in (\ref{metric0}) reads,
\begin{equation}
ds^2=
-\text{e}^{2\Phi(r)} dt^2+\frac{1}{g(r)} dr^2+r^2 d\Omega^2 \,,\label{metricWH}
\end{equation}
where $\Phi(r)$ is the so-called red-shift function and it depends on the radial coordinate only. 

A traversable wormhole is a space-time configuration characterized by a minimal radius or ``throat'' localized at $r=r_0\neq 0$ such that $g(r_0)=0$, while the function $\Phi(r)$ is finite and regular everywhere along the throat.
In the specific, the following conditions must be satisfied \cite{WHcond,WHcond2,WHcond3}:
\begin{itemize}
\item $\Phi (r)$ and $g (r)$ are regular and well defined for all $r \geq r_0$;
\item  $\Phi'_+(r_0)=\Phi'_-(r_0)$; 
\item $g(r_0)=0$ and $g(r)>0$ for all $ r \geq r_0$;
\item $g'_+(r_0)=g'_-(r_0)>0$.
\end{itemize}
The radial distance is given by
$
l(r)=\pm\int ^r_{r_0}\frac{d\tilde r}{g(\tilde r)} \,,
$
and is well defined everywhere \cite{Morris}. In particular, its minimal value is reached when $r=r_0$ where $l(r_0)=0$ and
its positive and negative values correspond to the lower and upper parts of the manyfold connected by the throat. Thus, the traveling time necessary to cross the wormhole between $l(r_1)<0$ and $l(r_2)>0$ is
$
\Delta t=\int_{-l(r_1)}^{l(r_2)}   \frac{d l}{v \text{e}^{\Phi(l)}} \,,
$
where $v=\frac{d l}{\text{e}^{\Phi(l)} dt}$ is the radial velocity of the traveler when passing a given radius $r$. The magnitude of $\Phi'(r)$ is associated with the tidal force experimented crossing the throat and big values of its modulus may make it difficult for an observer to complete the journey from one side to the other of the wormhole.

Given (\ref{metricWH}), the torsion scalar reads:
\begin{equation}
T=\frac{4 \left(\sqrt{\frac{1}{g(r)}}-1\right) g(r) \left(\sqrt{\frac{1}{g(r)}}-2 r \Phi '(r)-1\right)}{r^2}\,,
\end{equation}
and Eq. (\ref{T0}) corresponds to,
\begin{equation}\label{wwhh}
\frac{4 \left(\sqrt{\frac{1}{g(r)}}-1\right) g(r) \left(\sqrt{\frac{1}{g(r)}}-2 r \Phi '(r)-1\right)}{r^2}=T_0\,.
\end{equation}
It is now possible to make a general assumption for the metric function $g(r)$, namely
\begin{equation}\label{wwow}
g(r)=1-\frac{c_1}{r^z}\,,
\end{equation}
where $z > 0$ and $c_1>0$ are generic numbers chosen to be positive in order to satisfy the third and fourth above-mentioned conditions allowing the solution to be a wormhole. In this case, metric (\ref{metricWH}) may represent a wormhole whose throat is located at $r=c_1^{\frac{1}{z}}$. From Eq. (\ref{wwhh}) we get,
\begin{equation} 
\begin{split}
\Phi(r)=&c_2-\frac{1}{4 c_1 z (z+2) \sqrt{\frac{r^z}{r^z-c_1}} \left(r^z-c_1\right){}^{3/2}}\left( c_1 \frac{T_0}{2} z r^{z+2} \sqrt{1-\frac{r^z}{c_1}} \sqrt{r^z-c_1} \, _2F_1\left(\frac{1}{2},\frac{1}{2}+\frac{2}{z};\frac{3}{2}+\frac{2}{z};\frac{r^z}{c_1}\right) \right.
\\
& +\left(r^z-c_1\right) \left(\sqrt{r^z-c_1} \left(2 c_1 (z+2) \sqrt{\frac{r^z}{r^z-c_1}} \log
   \left(1-\frac{r^z}{c_1}\right) +\frac{T_0}{2} z r^{z+2} \left(1+\sqrt{\frac{r^z}{r^z-c_1}}\right) \right. \right. \\
   &\left. -2 c_1 (z+2) \sqrt{\frac{r^z}{r^z-c_1}} \log \left(1-c_1 r^{-z}\right)\right)+2 c_1 (z+2) r^{z/2} \log \left(1-\frac{r^{z/2}}{\sqrt{r^z-c_1}}\right)\\
   &\left. \left. -2 c_1 (z+2) r^{z/2} \log
   \left(1+\frac{r^{z/2}}{\sqrt{r^z-c_1}}\right)\right)  \right)\,,
\end{split}\label{Phi1}
\end{equation}
where $F_1(a,b;c;d)$ is the hypergeometric function.
Here, $c_2$ is a constant that only brings to a shift of the cosmological time, such that, without loss of generality, we can assume $c_2=0$. 

The expression above is quite involved and in what follows we will analyze the case $T_0=0$. After doing some calculations:
\begin{align}\label{wow}
\Phi(r)=\frac{1}{2} \log \left(\left(2-\frac{c_1}{r^z}+2\sqrt{1-\frac{c_1}{r^z}}\right)^{\frac{1}{z}}\right)\,,
\end{align}
or, equivalently, by reintroducing the metric function $h(r)$,
\begin{align}
h(r)= \left(2-\frac{c_1}{r^z}+2\sqrt{1-\frac{c_1}{r^z}}\right)^{\frac{1}{z}}\,.
\end{align}
If $z>0$ and $c_1>0$, equation (\ref{wow}) describes a regular well-defined function of all $r \geq r_0 = c_1^{1/z}$ and for $r=r_0$ we get $\Phi(r_0)=0$ while $\Phi'(r)>0$ when $r>r_0$. However, we should note that $\Phi'(r_0)$ diverges on the throat making it impossible to traverse the wormhole for extended bodies. Nonetheless, light can travel and carry information from one side to the other. According to the distinction reported in \cite{WHcond2} we should classify this wormhole as traversable in principle and not traversable in practice.

In order to find a wormhole that is traversable in practice, we would like to consider the following choice for $\Phi(r)$,
\begin{equation} \Phi(r)=\frac{1}{2} \log \left(\left(\frac{r}{\tilde r}\right)^z\right)\,,
\end{equation}
with $\tilde r$ a positive length parameter and $z>0$. 
In this case $h(r)=\left(\frac{r}{c_1}\right)^z$ and $\Phi'(r)>0$ for any value of $r$.
From Eq. (\ref{wwhh}) we get
\begin{equation}
    g(r)=\frac{2+2 z+z^2+r^2 \frac{T_0}{2} (z+1) \pm \sqrt{(z+2)^2 \left( r^2 T_0 (z+1)+z^2\right)}}{2 (z+1)^2}\,.
\end{equation}
By assuming $T_0>0$ and therefore by taking the solution with the minus sign we have
\begin{equation}
g(r_0)=0\quad \longleftrightarrow\quad
r_0= \sqrt{\frac{4}{T_0}}\,,
\end{equation}
and $g'(r_0)>0$.
We note that the radial value of the throat $r_0$ does not depend on the parameter $z$ and it is fixed by $T_0$.
It means that the class of models satisfying (\ref{Ansatz}) for some value of $T_0\neq 0$ admits traversable wormhole solutions whose throat is determined by the model parameter $T_0$ only.

\section{Rotation curves of galaxies\label{rotation}}

The analysis of the rotation curves of galaxies provides another playground where SSS metrics have useful applications.
Galaxies are characterized by two ingredients: the visible baryonic matter forming a rotating disk, and some electromagnetic invisible matter, the so-called dark matter, forming a
spherical halo that encloses the baryonic matter disk. The dark matter halo is required to explain the observed flattening of the rotation curve of the baryonic matter at a large distance from the galactic center (see Refs.~\cite{Salucci:2007et, gal in univ} for a review on such topics). Phenomenologically, they result in a Newtonian potential of the form ~\cite{Riegert:1984zz, Mannheim:1988dj, Mannheim:2005bfa, Mannheim:2010ti, Mannheim:2010xw, OBrien:2011vks, Capozziello:2006ph, Salucci:2014oka, Salucci:2013rmp, Hashim:2014mka},
\begin{equation}
 g_{tt}(r)=-h(r) = - (1 + 2\varphi_{tot}(r)) = - (1 + 2\varphi_{BM}(r) + 2\varphi_{DM}(r)) = -(1-\frac{c_0}{r}+ c_1 r) \,,
\end{equation}
where $c_0$, $c_1$ are positive dimensional constants, $\varphi_{tot}(r)$ is the total potential and $\varphi_{BM}(r)$ and $\varphi_{DM}(r)$ are its baryonic and dark matter components, respectively. The baryonic matter potential is in the form of the classical Newtonian potential scaling according to the radial distance, while the dark matter potential is linear with respect to the radial distance, in order to reproduce the observed flattening of the rotation curves.

By neglecting the behavior for small values of $r$, it is possible to derive an interpolating form as, \cite{Roberts:2002ei}
\begin{equation}
    h(r)=1+c_2 \log{r}\,,
    \label{log}
\end{equation}
where $c_2$ is a positive constant. On the other hand,  for large values of $r$ we get,
\begin{equation}
    h(r)=1+c_1 r\,.\label{r}
\end{equation}
In the framework of a modified theory of gravity, the effects of dark matter can be played by modification of gravity without invoking any exotic form of matter. In this section, we will analyze exact SSS solutions in the form of (\ref{metric0}) and whose metric function $h(r)$ is given by (\ref{log}) or (\ref{r}) and which lead to a constant torsion scalar $T_0$, namely they are solutions of the class of models for which Eq. (\ref{Ansatz}) holds true.

First of all, we remind that, when $T_0=0$, the picture is significantly simplified. In this case, as we already observed in \S \ref{ModelAn}, a solution of Eq. (\ref{T0}) is given by $g(r)=1$. It means that the space-time described by
\begin{equation}
ds^2=-h(r)dt^2+dr^2+r^2d\Omega^2\,,
\end{equation}
with $h(r)$ given by (\ref{log}) or (\ref{r}) and which reproduces the phenomenology of dark matter is an exact solution of all the modified teleparallel gravity models satisfying Eq. (\ref{Ansatz}) with $T_0=0$.
Beyond this choice, other possibilities are allowed for the metric function $g(r)$ according to (\ref{log}) or (\ref{r}). In the first case, when $T_0=0$, we have,
\begin{equation}
    g(r)=\frac{c_2{}^2 \log ^2(r)}{(c_2 \log (r)+1+c_2){}^2}+\frac{2 c_2 \log (r)}{(c_2 \log (r)+1+c_2){}^2}+\frac{1}{(c_2 \log (r)+1+c_2){}^2}\,,
\end{equation}
which tends to 1 for large values of $r$. 

In the second case, one finds a general solution for $T_0\neq 0$, namely,
\begin{align}
   g(r)=&\frac{c_1{}^2 r^4 T_0}{2(1+2 c_1 r){}^2}+\frac{3 c_1 r^3 T_0}{4 (1+2 c_1 r){}^2}+\frac{r^2 T_0}{4 (1+2 c_1 r){}^2}+\frac{5 c_1{}^2 r^2}{2 (1+2 c_1 r){}^2}+\frac{3 c_1 r}{(1+2 c_1 r){}^2}\\
   &+\frac{1}{(1+2 c_1 r)^2 } \pm \frac{\sqrt{r^2 (2+3 c_1 r){}^2 \left(T_0 \left(2 c_1{}^2 r^2+3 c_1 r+1\right) +c_1{}^2\right)}}{2 (1+2 c_1 r){}^2}\,,
\end{align}
which for $T_0=0$ reduces to 
\begin{align}
g(r)=\frac{5 c_1{}^2 r^2+6 c_1 r+2 \pm \sqrt{c_1{}^2 r^2 (2+3 c_1 r){}^2}}{2 (1+2 c_1 r){}^2}\,,
\end{align}
such that by taking the plus sign for large values of $r$ we still have $g(r) \rightarrow 1$.

\section{Conclusions}

In this paper, we analytically studied exact solutions of $f(T)$-gravity having a static and spherical symmetry. 

In TEGR the gravitational lagrangian corresponds to the torsion scalar $T$, while $f(T)$-gravity is an extension of the theory where the gravitational lagrangian is a generic function of the torsion scalar. 
The SSS solutions we found are characterized by a constant (eventually
vanishing) torsion scalar and belong to models satisfying a general Ansatz for which the equations of motion are trivially satisfied. SSS solutions may describe compact objects such as black holes and wormholes as well as take into account the anomalous rotation curve typical of spiral galaxies.

We investigated several metrics describing black hole solutions that differ from the Schwarz-schild metric thus opening the possibility of comparing the results with Solar System tests, gravitational waves coming from black hole mergers, and black hole shadow to constrain the parameter space.

The wormhole solutions we found have configurations leading to wormholes that are traversable in principle and in practice.
Such solutions display a conceptual advantage, in fact, they are not in need of odd matter sources violating the null energy condition and preventing the throat from pinching off. This exotic matter source is needed in the context of GR and can be made of additional new fields or by the anomalous trace of standard fields \cite{Calza:2022szy}. For what concerns the wormhole solutions we proposed in this paper, this unpleasant role is played by the gravitational sector which is modified and no additional exotic matter is invoked.

The last offspring of our Ansatz consists of solutions capable of mimicking the influence of the dark matter halo in the rotational behavior of spiral galaxies. As for the other cases, there exists the potentiality of exploiting the rotation curves of a large number of galaxies for fixing the parameters.


\section*{Acknowledgements}
This work was supported by national funds from FCT, I.P., within the projects UIDB/04564/2020, UIDP/04564/2020 and the FCT-CERN project CERN/FIS-PAR/0027/2021. \newline
M.C. is also supported by the FCT doctoral grant SFRH/BD/146700/2019.

\end{document}